\documentclass[fleqn,10pt]{wlscirep}
\usepackage[utf8]{inputenc}
\usepackage[T1]{fontenc}
\usepackage{bm}
    \title{Chiral phase memory of twisted light through multiple scattering}

\author[1,*]{Igor Meglinski}
\author[2]{Anton Sdobnov}
\author[2]{Alexander Bykov}
\affil[1]{Aston Institute of Photonic Technologies, College of Engineering and Physical Sciences, Aston University, Birmingham, B4 7ET, UK}
\affil[2]{Optoelectronics and Measurement Techniques, University of Oulu, P.O. Box 4500, Oulu, FI-90014, Finland}

\affil[*]{Correspondence: i.meglinski@aston.ac.uk}

\begin{abstract}
Chiroptical signals, optical responses sensitive to molecular handedness, are rapidly suppressed by multiple scattering, fundamentally limiting their use in turbid media. Here we show that coupling molecular chirality to the topological structure of twisted light generates a protected phase observable that survives strong scattering. When Laguerre–Gaussian beams carrying orbital angular momentum propagate through chiral media, spin–orbit interaction converts circular birefringence into an azimuthal rotation of the helical wavefront. Remarkably, this chiral phase memory persists at scattering strengths that fully depolarize conventional beams, with the rotation magnitude preserved quantitatively between transparent solutions and strongly scattering tissue. The sign of the azimuthal rotation encodes molecular handedness: opposite enantiomers produce mirror-symmetric phase maps even after multiple scattering. Differential measurements between conjugate topological charges isolate the chiral contribution while cancelling achiral background, enabling the resolution of refractive-index changes of order $10^{-6}$. These results establish topological phase observables as robust carriers of weak chiral light–matter interactions in complex media, opening new routes for chiroptical spectroscopy and sensing beyond the ballistic-photon regime.
\end{abstract}
\begin{document}
\flushbottom
\maketitle
\thispagestyle{empty}


\section*{Introduction}
Optical memory effects represent a fundamental class of phenomena in which electromagnetic waves retain partial correlations with their initial spatial, angular, temporal, or polarization characteristics after propagating through complex disordered or structured media, contrary to classical expectations of complete randomization through multiple scattering. The concept of optical memory emerged from early investigations into wave propagation through disordered media, fundamentally challenging the classical view that multiple scattering completely randomizes electromagnetic waves. Theoretical work by Kaveh~\textit{et al.} in 1986 laid the foundation for understanding enhanced backscattering phenomena~\cite{Kaveh}, which was soon followed by experimental demonstrations by Mark~\textit{et al.} in 1988, revealing coherent interference effects in the exact back-scattering direction~\cite{Mark}. The same year marked a pivotal moment in the field: the optical memory effect was first predicted theoretically and verified experimentally, demonstrating that spatial information of incident waves is not entirely lost but partially preserved through transmission or reflection from opaque materials, exhibiting linear angular correlations and exponential decay with increasing deviation~\cite{Feng,Freund}. This discovery challenged the prevailing assumption that multiple scattering events would completely scramble wave correlations, revealing instead that deterministic interference processes maintain measurable correlations between input and output wavefronts. Further experimental and theoretical studies have shown that multiple scattering preserves more information than previously assumed. Beyond spatial correlations, polarization memory has been observed in turbid medium~\cite{MacKintosh}, along with weak localization and temporal coherence retention on ultrafast timescales~\cite{Yoo:89}, and interference effects arising from time-reversal symmetry in coherent backscattering~\cite{Akkermans,Freund:94}. These results established a broader physical framework for understanding optical memory phenomena in disordered systems~\cite{Kravtsov:93}.

Subsequent systematic exploration of optical memory effects led to the identification of distinct correlation types and their practical implications. Angular memory was investigated for target detection in scattering media~\cite{Chan,Zhang:98}, while double-passage correlations through random phase screens were demonstrated experimentally~\cite{LinGu}. These developments underscored the robustness of coherence and polarization-related memory effects across a wide range of scattering environments, from biological tissues to cold atomic ensembles. The persistence of circular polarization in turbid media~\cite{Kim:06}, and the controlled manipulation of coherent backscattering in cold atomic gases~\cite{Labeyrie}, demonstrated that multiple scattering does not erase structured information, but redistributes it in predictable and recoverable ways. This growing understanding laid the groundwork for exploiting more complex degrees of freedom of light in disordered systems, ultimately opening new directions for high-resolution sensing, imaging, and information retrieval~\cite{KUPRIYANOV20171}.

Building on earlier discoveries, researchers uncovered new classes of correlations in anisotropically scattering biological media~\cite{Judkewitz}, while extended angular memory ranges were demonstrated experimentally in tissue-mimicking phantoms and biological tissues as a result of pronounced forward scattering~\cite{Schott}. Phase-space measurements enable depth-resolved imaging of hidden objects~\cite{Takasaki}. These findings reinforced the understanding that biological tissues, despite their structural complexity, can sustain long-range spatial and angular coherence, paving the way for advanced optical imaging and sensing techniques in highly scattering environments. 

Thus, the field entered its modern phase with the introduction of the generalized optical memory effect framework, which unified spatial and angular correlations across diverse scattering regimes~\cite{Osnabrugge}. This formalism, provided a comprehensive theoretical and computational basis for understanding wave correlations beyond the classical tilt-tilt paradigm. Building on this, recent advances in wavefront shaping have enabled active control over memory effects in complex media: Yılmaz and colleagues demonstrated that angular correlations can be arbitrarily modified through engineered transmission channels and tailored input fields~\cite{Yilmaz}. 

Wavefront shaping techniques have emerged as a promising approach for imaging through opaque media and correcting optical aberrations~\cite{Zhang,Katz}. These methods exploit the optical memory effect, which allows for the manipulation of scattered light~\cite{WangX}. These developments mark a shift from passive observation of memory effects to their programmable exploitation, opening new frontiers for imaging, metrology, and light–matter interaction in disordered systems~\cite{Park,Yu}.

In addition, recent research has focused on the propagation of structured light carrying orbital angular momentum (OAM)~\cite{Padgett-Allen} in turbid, tissue-like scattering media~\cite{Wang:16,Ntziachristos}, exploring their potential to retain structured information despite multiple scattering events~\cite{Viola,Daryl,Fu:24}. Unlike scalar field correlations or polarization states that degrade rapidly in biological tissues~\cite{TuchinJBO}, the topological charge of twisted light exhibits remarkable resilience, maintaining measurable phase characteristics despite complex scattering~\cite{Khanom}. This phase memory effect, first demonstrated in~\cite{Meglinski}, reveals that structured light-carrying OAM retains quantum-like topological protection even in highly disordered media, as highlighted in~\cite{Novikova}. This phenomenon fundamentally differs from classical optical memory effects, which typically preserve only angular or spatial correlations through statistical averaging. The OAM phase memory arises from the topological nature of the helical wavefront, a property that cannot be gradually degraded but only discretely changed. This topological protection enables optical measurements through tissue with unprecedented sensitivity, as phase shifts induced by minute refractive index changes remain detectable despite multiple scattering.

Here, we demonstrate that molecular chirality couples to the topological structure of OAM light through spin-orbit interaction, generating a chiral phase observable that survives strong multiple scattering. This chiral-topology coupling converts circular birefringence into an azimuthal rotation of the helical wavefront, a topologically protected quantity that remains quantitatively accessible even when conventional chiroptical signals are destroyed. We validate this principle using glucose enantiomers as model chiral molecules, demonstrating that handedness information persists through scattering strengths far exceeding polarization survival limits.

\section*{Results}

\subsection*{Chiral phase preservation across scattering regimes}

\begin{figure}[h!]
\begin{center}
    \centering
    \caption{\textbf{Chiral phase memory persists across scattering regimes}. \\
    \includegraphics[width=1.00\linewidth]{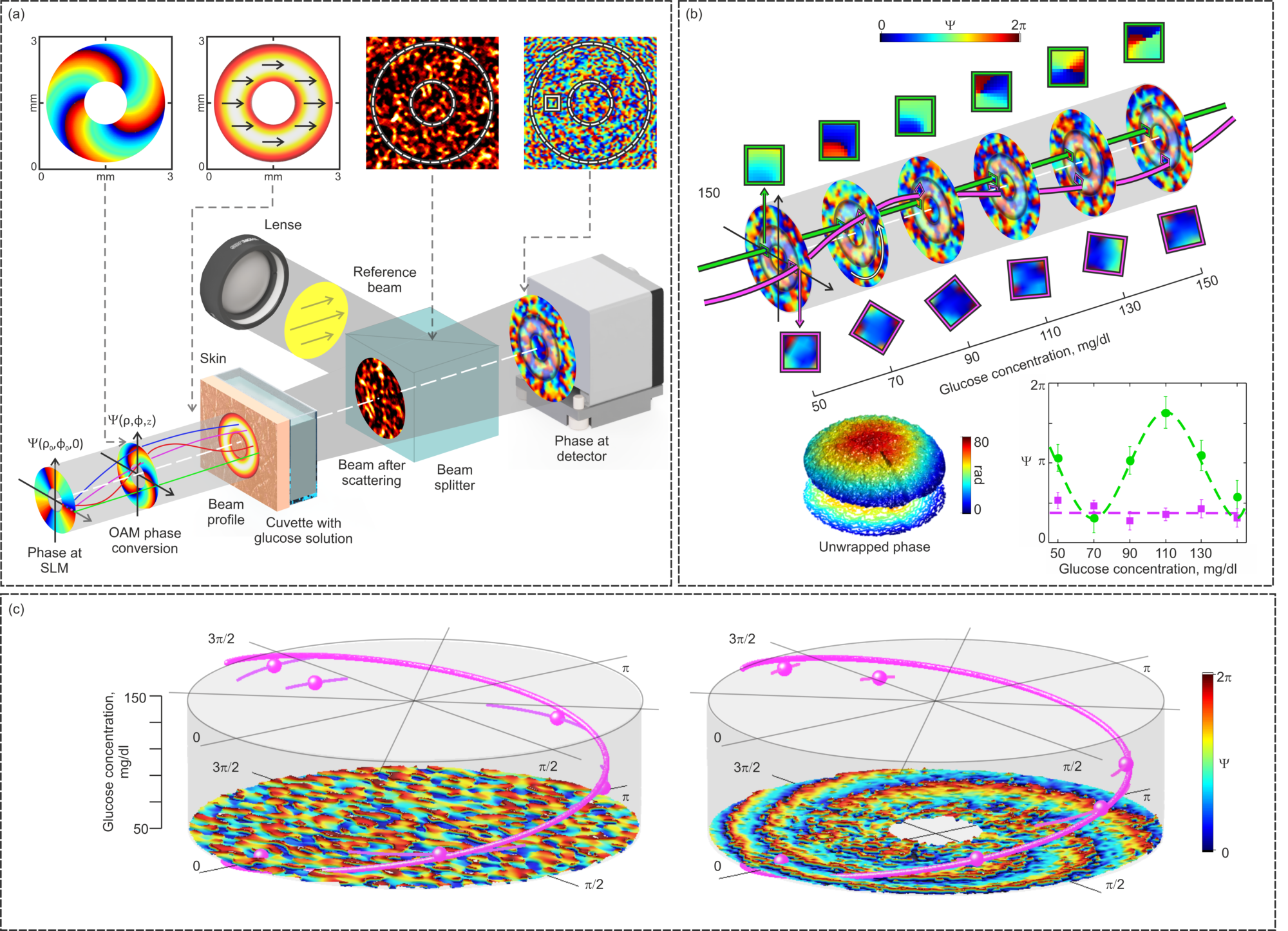}\\
        (a) Schematic representation of the central part of the Mach–Zehnder interferometer  experimental system~\cite{Meglinski}: the LG beam ($\ell = 3, p = 0$), carrying OAM imparted by a spatial light modulator (SLM), traverses \textit{ex vivo} skin sample and cuvette containing glucose solution in water; (b) Interference between the transmitted LG beam and a reference plane wave (initially an expanded Gaussian beam), captured by camera for phase retrieval analysis, for the glucose concentration range of $C \in [50,150]~mg/dl$; (c) OAM-related phase shift versus glucose concentration $C$ for: (left) \textit{ex vivo} porcine skin tissue (strongly scattering, $z/l^* \approx 10.0$) placed before a cuvette containing glucose solution; (right) nearly transparent tissue phantom (weakly scattering, $z/l^* = 0.1$). Linear fits demonstrate that the rate of OAM twist preserves the quantitative relationship despite varying scattering strength. Error bars represent standard deviation from $n = 10$ measurements.} 
        \label{fig:glucose1} 
\end{center}
\end{figure}

We first establish that chiral information encoded in twisted light survives multiple scattering by measuring azimuthal phase rotations through media of varying turbidity. Figure~\ref{fig:glucose1}-(a) shows the experimental configuration: Laguerre-Gaussian (LG) beams carrying OAM traverse scattering samples before encountering chiral solutions, with the transmitted field interfering with a reference beam for phase retrieval. Despite strong speckle formation in the scattered field (see Fig.~\ref{fig:glucose1}-(b)), the azimuthal phase shift $\Psi_{OAM}$ remains well-defined and responds quantitatively to the chiral medium. We use aqueous glucose solutions as a model system, exploiting the well-characterized circular birefringence of this chiral molecule to provide a calibrated test of phase preservation. The key observation is presented in Fig.~\ref{fig:glucose1}-(c): the slope relating phase shift to concentration is preserved across scattering strengths spanning two orders of magnitude—from nearly transparent phantoms ($z/l^* = 0.1$) to strongly scattering tissue ($z/l^* \approx 10$). This invariance demonstrates that the azimuthal phase gradient, rather than the absolute optical phase, acts as the protected observable. Here, phase memory functions not merely as a passive correlation effect but as an active transduction mechanism, converting molecular chirality into a topologically robust phase observable that survives multiple scattering. 
Additional measurements demonstrating the robustness of this behavior across different scattering regimes, opposite molecular handedness, and multiple OAM topological charges ($\ell = \pm 3$ and $\pm 5$) are presented in Supplementary Figure.

Quantitatively, the observed total glucose-induced OAM-related phase shift can be defined as~\cite{Allen,Meglinski,Berry2008}:
\begin{equation}
\Psi_{meas}(C, \ell) =  k \, \Delta n_{eff}(C) \, L_{spiral}(\rho,\phi,z) \, \ell,
\label{eq:phaseshift}
\end{equation}
where $\ell$ is the topological charge of the LG beam, $k$ is the wavenumber $(k = 2\pi / \lambda)$ with the wavelength of light in vacuum $\lambda$; $L_{spiral}$ denotes the effective spiral-like photon trajectory length associated with OAM-carrying LG beam propagation~\cite{Meglinski,Berry2008,Bliokh2013}, defined as a function of cylindrical coordinates: radial ($\rho$), azimuthal ($\phi$), and longitudinal ($z$), which naturally describe the helical phase structure and propagation geometry of OAM-carrying beams in the medium. In turbid tissue-like scattering medium, such as \textit{ex vivo} porcine skin, the effective refractive-index increment $\Delta n_{eff}(C)$ encompasses both achiral and chiral contributions to the optical response.

The achiral component arises from bulk refractive index changes, optical path elongation due to multiple scattering, and concentration-dependent polarizability of glucose molecules in solution~\cite{TuchinJBO}. These contributions produce a global phase delay that is invariant under reversal of molecular handedness and under sign inversion of the OAM topological charge $(\ell \to -\ell)$. This symmetry follows from the scalar nature of bulk refractive index variations: random multiple scattering does not distinguish between opposite helical phase gradients $\pm \ell$, whereas chiral light–matter interactions couple differentially to the handedness of both the molecular structure and the optical vortex~\cite{Cameron_2012,Forbes:18}.

The chiral component of the optical response arises from circular birefringence, defined as
\begin{equation}
\Delta n_{CB}(C) = n_L(C) - n_R(C), 
\label{eq:deltan}
\end{equation}
which produces concentration-dependent optical rotation in chiral media~\cite{Barron_2004}. Thus, the effective refractive index increment decomposes as
\begin{equation}
\Delta n_{eff}(C, \ell) = \Delta n_{achiral}(C) + \text{sign}(\ell) \cdot \Delta n_{CB}(C), 
\label{eq:deltan2}
\end{equation}
where the achiral term is independent of topological charge while the chiral term changes sign with $\ell$. In beams carrying OAM, this chiral phase delay is transduced via spin–orbit interaction into an azimuthal rotation of the helical wavefront~\cite{Bliokh,Bliokh_2017}. Owing to the OAM phase memory effect~\cite{Meglinski}, the ensemble-averaged helical phase structure, and therefore the chiral OAM-dependent phase shift, remains accessible even under strong multiple scattering, whereas non-chiral phase contributions appear as a common-mode background that cancels in differential measurements between conjugate topological charges.

\subsection*{Differential detection and topological charge optimization}

This decomposition motivates our experimental strategy for isolating the glucose-specific signal. By comparing phase shifts for $\ell = +5$ and $\ell = -5$ beams through identical glucose samples, the achiral background cancels:
\begin{equation}
\Delta\Psi_{diff} = \Psi_{meas}(\ell = +5) - \Psi_{meas}(\ell = -5) = 2k \, \Delta n_{CB}(C) \, L_{spiral} \cdot |\ell|, 
\label{eq:diff}
\end{equation}
isolating the glucose-specific chiral signal with doubled sensitivity. Furthermore, comparing responses to $D(+)$ and $L(-)$ glucose enantiomers provides an additional consistency check, as opposite molecular handedness produces opposite signs of $\Delta n_{CB}$ while maintaining identical achiral contributions.

The enhanced sensitivity originates from the spiral energy-flow geometry intrinsic to OAM beams, which increases the effective interaction length for chiral phase accumulation~\cite{Berry2008,Bliokh2013}.
Geometric and achiral phase contributions affect both topological charges identically and therefore cancel in differential measurements, whereas the glucose-specific chiral term reverses sign with $\ell$ and is retained. The full phase decomposition used for this separation is given in the Methods. However, the $|\ell|$ dependence in (\ref{eq:diff}) encodes additional geometric information that merits deeper examination.

The relationship between topological charge and detection sensitivity reflects the intrinsic geometry of energy flow in twisted light. In LG beams, the Poynting vector traces helical trajectories whose pitch depends inversely on $|\ell|$~\cite{Berry2008,Bliokh2013}: lower-order modes ($\ell = 3$) generate loosely wound helices with large pitch, while higher-order modes ($\ell = 5$) produce tightly wound trajectories executing multiple revolutions per Rayleigh length~\cite{Allen1999-uc}. This geometric distinction creates complementary metrological regimes.

For loosely wound trajectories, the azimuthal phase gradient $\partial \Psi/\partial \phi = -\ell$ varies slowly around the circumference of the beam~\cite{Allen}. Small perturbations to the refractive index therefore produce proportionally large distortions of this gentle phase structure analogous to grazing-incidence interferometry, where shallow angles amplify sensitivity to path-length differences. Lower topological charges thus provide enhanced fractional sensitivity to minute refractive index variations, optimal for detecting weak chiroptical signals in transparent media.

Conversely, tightly wound trajectories in higher-order modes accumulate greater absolute phase through their extended helical path length, as captured by the $|\ell|$ factor in (\ref{eq:diff})~\cite{Berry2008}. This geometry favours differential measurements in scattering media, where the larger absolute phase shift between conjugate charges ($+\ell$ and $-\ell$) improves signal-to-noise ratio despite the reduced fractional sensitivity per unit refractive index change.

This geometry-sensitivity relationship reveals a fundamental design principle for OAM-based chiroptical detection: the topological charge functions as a tunable parameter that balances differential sensitivity against absolute signal strength. Our experimental strategy exploits this trade-off directly, employing $\ell = \pm 3$ for initial calibration in transparent solutions where fractional sensitivity dominates, and $\ell = \pm 5$ for differential chiral detection through scattering tissue where absolute phase contrast determines measurement fidelity. 

\subsection*{Enantiomer-selective phase rotation through tissue}

With this geometric framework established, we now examine the phase structure experimentally. Figure~\ref{Phase} illustrates schematically the geometric phase decomposition (\ref{Psi}). The intrinsic geometric terms establish the baseline helical phase structure, while the glucose-induced chiral contribution manifests as a coherent azimuthal rotation of this structure. 
Equation (\ref{eq:diff}) shows that differential comparison of conjugate OAM states removes all geometric and achiral contributions, leaving a purely chiral phase term that is doubled in amplitude and directly proportional to the glucose-induced circular birefringence.

\begin{figure}[h!]
\begin{center}
    \centering
        \caption{\textbf{Phase decomposition for OAM-based chiral detection}. \\
       \includegraphics[width=1\linewidth]{Phase_formation4.png}\\
        Individual phase contributions to total LG beam phase $\Psi(\rho,\phi,z;C,\ell)$ at the detection plane for $\ell = 5$, respectively: Radial wavefront curvature $-k \rho^2 z/2(z^2 + z^2_R)$, arising from beam divergence; Azimuthal helical phase $-\ell \phi$ (carrying topological charge); Longitudinal propagation phase $-kz$; the Gouy phase $G(z) =(2p + |\ell| + 1)arctan(z/z_R)$; Complete geometric phase including radial curvature, helical, longitudinal propagation and Gouy phases at the cuvette entrance ($z = 70~cm$); and Scattering-free medium-induced phase accumulation $\Psi_{\text{medium}}(C,\ell)$, manifesting as coherent rotation of the helical structure by angle $\theta_{\text{OAM}} \propto \Delta n_{\text{CB}}(C)$ relative to reference concentration ($C = 50~mg/dl$). In transparent (non-scattering) media, the achiral contribution $\Psi_{\text{achiral}} = 0$, and the observed phase shift arises purely from glucose-induced circular birefringence. Color scale: phase from $\pi$ to $2\pi$.} 
        \label{Phase}    
\end{center}
\end{figure}

\begin{figure}[htbp]
\begin{center}
    \centering
        \caption{\textbf{Chiral phase discrimination between glucose enantiomers through multiple scattering}. \\
       \includegraphics[width=0.75\linewidth]{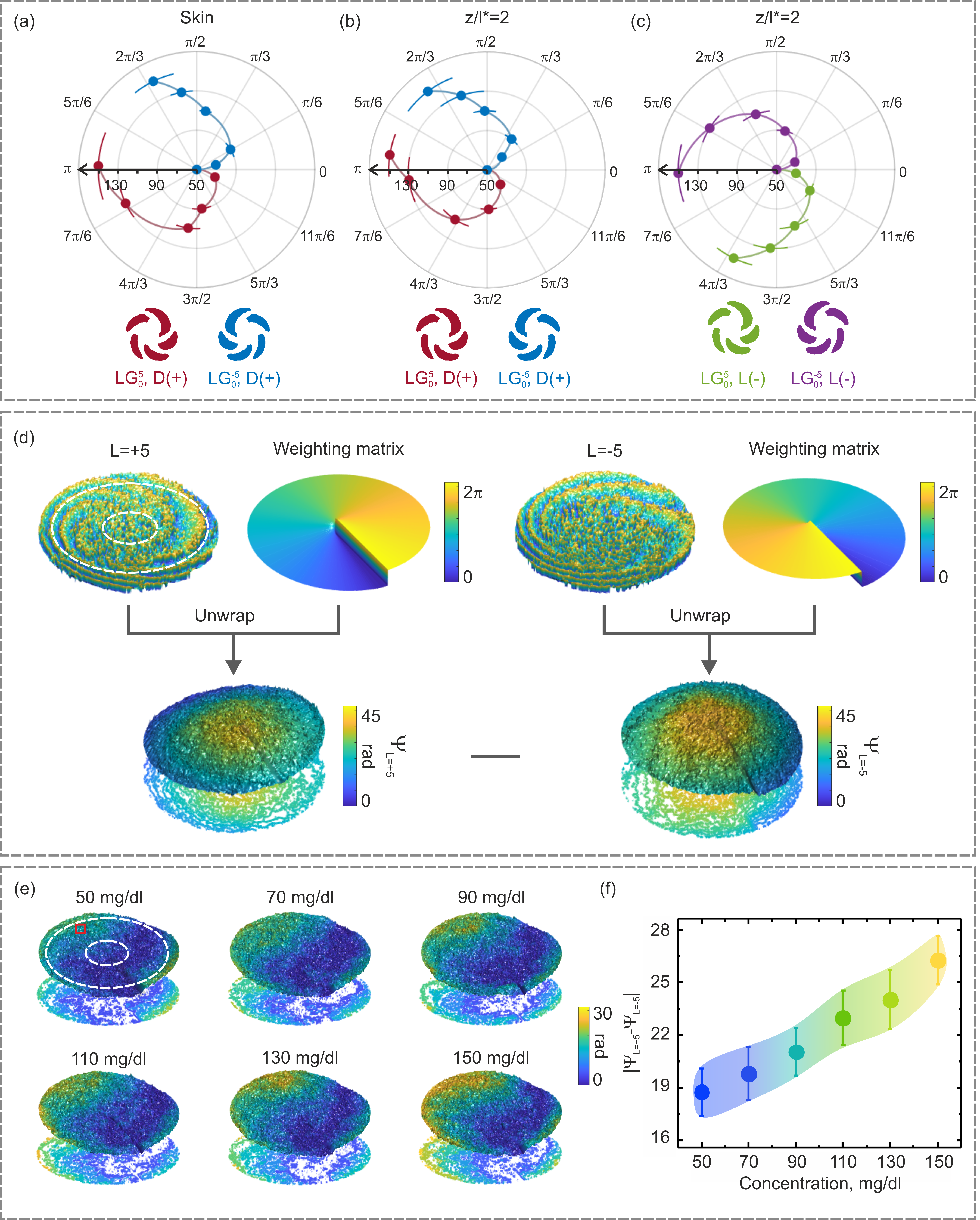}\\
       Polar plots of azimuthal rotation $\theta_{\text{OAM}}$ extracted from unwrapped phase maps $\Psi(\rho,\phi)$ at fixed radius $\rho$: (a) \textit{ex vivo} porcine skin ($z/l^* \approx 10$) and (b) tissue phantom ($z/l^* = 2$) for $D(+)$-glucose; (c) tissue phantom for L(-)-glucose, measured with opposite topological charges; $\ell = +5$ and $\ell = -5$. (d) Phase retrieval workflow: wrapped vortex phase and corresponding weighting matrices (top row), and resulting unwrapped phase maps (bottom row), obtained by off-axis vortex interferometry (referenced to $C=50~mg/dl$). (e) Differential phase maps $\Psi_{\ell=+5}-\Psi_{\ell=-5}$ through \textit{ex vivo} porcine skin for glucose concentrations $50 - 150~mg/dl$. (f) Differential phase $|\Psi_{\ell = +5} - \Psi_{\ell = -5}|$ versus glucose concentration, isolating the chiral glucose-induced contribution. Symbols show mean over $n = 10$ independent measurements; error bars indicate standard deviation.} 
        \label{Gluc}    
\end{center}
\end{figure}

Figure~\ref{Gluc} directly probes whether the azimuthal phase of OAM light retains quantitative information after multiple scattering. Despite strong speckle-induced phase scrambling, the ensemble-averaged azimuthal phase gradient remains well defined and exhibits a reproducible, concentration-dependent rotation. The differential phase shift, $\Delta \Psi_{diff}$, between opposite topological charges ($\ell = +5$ and $\ell = -5$), maintains a linear relationship with glucose concentration even in the presence of strongly scattering media. Crucially, this linear relationship is preserved with near-identical slope through both \textit{ex vivo} porcine skin ($z/l^* \approx 10$) and the low-scattering tissue phantom ($z/l^* = 2$), spanning an order of magnitude in scattering strength. 
The invariance of the slope across an order-of-magnitude change in scattering strength demonstrates that the azimuthal phase gradient, rather than the absolute optical phase, acts as the preserved quantity. This satisfies the operational definition of phase memory: a measurable phase observable that remains quantitatively linked to the input perturbation despite multiple scattering. Figure~\ref{Gluc} therefore establishes azimuthal phase rotation as the experimentally accessible carrier of OAM phase memory in scattering media.

The physical origin of this remarkable preservation can be understood through the theoretical framework developed in (\ref{eq:phaseshift})–(\ref{eq:diff}). The achiral contributions, arising from bulk refractive index modulation, scattering-induced path length variations, and depolarization, affect both topological charges identically and therefore vanish upon subtraction. Only the circular birefringence term, which reverses sign under $\ell \to -\ell$, survives and is doubled in amplitude.

Critically, the glucose term modifies the azimuthal phase relationship, producing a measurable rotation of the helical structure proportional to concentration. Experimentally, this cancellation is visualized in Fig.~\ref{Gluc}-(e): although the absolute phase maps vary strongly with scattering and concentration, their difference reveals a smooth, monotonic azimuthal rotation. The persistence of this differential signal indicates that scattering acts as a common-mode perturbation, while the chiral phase term is preserved as a relative phase invariant.

Although multiple scattering randomizes the absolute phase at fixed spatial coordinates ($\rho,\phi$), evident in the complex speckle patterns, the azimuthal phase gradient integrated along spiral-like trajectory preserves the helical structure (see Fig.~\ref{fig:glucose1}-(b)), enabling quantitative phase retrieval despite tissue turbidity. Multiple scattering destroys the spatially local phase but does not erase the azimuthal phase gradient associated with OAM. The experimentally observed equivalence of $\theta_{OAM}$ measured through \textit{ex vivo} skin ($z/l^* \approx 10$) and weakly scattering phantoms confirms that the helical phase structure survives as an ensemble property. This demonstrates that OAM phase memory is not a ballistic-photon effect but a robust statistical invariant.

The observed OAM-based phase shift $\Psi_{OAM}$ originates from the interplay between glucose-induced optical activity and the intrinsic spiral geometry of OAM light propagation. As glucose concentration increases within the physiological range ($50 - 150~mg/dl$), the enhanced optical activity of the medium induces cumulative polarization rotation along the helical photon trajectories of the LG beam. Critically, the characteristic spiral trajectories of OAM LG beams traverse different optical path lengths through the glucose-containing medium, with trajectories near the beam periphery following helical paths longer than those near the center. Owing to the rotational symmetry of the LG mode, such polarization-mediated changes preserve the topological structure while producing a measurable azimuthal phase shift. The collective effect across all spiral trajectories therefore appears as a coherent twist of the OAM wavefront, experimentally observed as $\Psi_{OAM}$. The measured phase evolution constitutes a direct manifestation of the OAM phase-memory effect~\cite{Meglinski}, but with a crucial distinction: whereas conventional phase memory in scattering media arises from stochastic preservation of topological structure through random photon deflections, here the modulation is deterministically imposed by glucose-induced optical activity. This OAM-based phase memory approach demonstrates high sensitivity to minute refractive index variations, with phase shifts resolvable down to $\Delta n \approx 10^{-6}$. These results are in good agreement with alternative studies employing twisted OAM light in various interferometer configurations~\cite{Review,Verma,Fedoruk,Munjal}, confirming the reliability of OAM-based phase detection for resolving refractive index changes in transparent media with high sensitivity.

When the LG beam propagates through strongly scattering tissue, individual ballistic trajectories are destroyed and the output exhibits complex speckle patterns (Fig.~\ref{fig:glucose1}-(b)). However, the OAM phase memory effect~\cite{Meglinski} preserves the global topological structure characterized by $\ell$. The glucose-induced circular birefringence $\Delta n_{CB} = n_L(C) - n_R(C)$ produces a collective rotation $\theta_{OAM}(C)$ of this preserved helical phase, a statistical property of the scattered photon ensemble rather than individual paths. 

The intrinsic chirality of glucose molecules drives a concentration-dependent optical rotation that systematically twists the OAM wavefront (see Fig.~\ref{fig:glucose1}-(c)), converting a passive statistical effect into an actively tunable sensing principle. Each incremental rise in glucose concentration translates into a proportional azimuthal rotation of the helical phase front, yielding a linear calibratable response. Experimentally, the glucose response manifests as a concentration-dependent azimuthal rotation $\theta_{OAM}$ of the preserved helical phase, whose magnitude scales linearly with glucose concentration.

In addition, the enantiomer selectivity arises from spin-orbit interaction (Fig.~\ref{Gluc}). 

The chirality-dependent nature of the interaction is directly evidenced by the opposite azimuthal phase rotations observed for $D(+)$- and $L(-)$-glucose at identical concentrations (Fig.~\ref{Gluc}-(c)). While the magnitude of the OAM-induced rotation scales linearly with concentration in both cases, its sign reverses with molecular handedness, yielding mirror-symmetric phase maps. This enantiomer-specific response, obtained under identical optical and scattering conditions, confirms that the measured azimuthal phase shift originates from circular birefringence rather than achiral refractive-index effects.

Remarkably, this chirality-dependent signature persists even through strongly scattering media, demonstrating that the topological protection of the OAM phase memory effect preserves not only the magnitude but also the sign of the glucose-induced optical activity. This preservation of enantiomer-specific information through turbid tissue represents a significant advantage for biomedical sensing, as it provides an additional layer of molecular specificity beyond simple concentration measurements. The generality of this enantiomer-specific response, including its preservation under both weakly and strongly scattering conditions and for different topological charges, is further confirmed by independent datasets presented in Supplementary Figure. 

The spiral trajectory formalism~\cite{Meglinski} makes specific, testable predictions about how molecular chirality couples to beam topology. We validate this mechanism through systematic measurements with glucose enantiomers and opposite topological charges. Figure~\ref{Gluc} presents compelling experimental evidence for the circular birefringence mechanism: $D(+)$ and $L(-)$ glucose enantiomers at identical concentrations produce opposite azimuthal rotations of the OAM phase across low and multiple scattering media. This enantiomer-specific response unambiguously demonstrates that the measured phase shift originates from molecular chirality rather than simple refractive index changes, as only the distinct handedness of $D(+)$ (dextrorotatory) versus $L(-)$ (levorotatory) glucose can induce opposite signs of circular birefringence $\Delta n_{CB} = n_L - n_R$. The observed sign reversal directly validates our model: the measured phase shift $\Psi_{meas}$ depends linearly on $\Delta n_{CB}$, with $D(+)$ producing $\Delta n_{CB} > 0$ (positive rotation) and $L(-)$ producing $\Delta n_{CB} < 0$ (negative rotation), as confirmed experimentally in Fig.~\ref{Gluc}. Crucially, both the direction and magnitude of $\theta_{OAM}$ are preserved through \textit{ex vivo} porcine skin ($z/l^* \approx 10$), demonstrating that OAM phase memory retains handedness information even under strong multiple scattering, a fundamental distinction from conventional polarimetry, where scattering rapidly suppresses enantiomer sensitivity.

The phase shift $\Psi_{meas}(C)$ exhibits opposite signs for conjugate topological charges: $D(+)$ glucose induces clockwise twist for $\ell = +5$ but negative for $\ell = -5$, while $L(-)$ glucose produces the inverse pattern (see Fig.~\ref{Gluc}). This topological charge-dependent response arises from spin-orbit coupling: the handedness of the helical wavefront determines whether glucose-induced circular birefringence adds to or subtracts from the accumulated phase. This chirality-topology coupling doubles the measurement sensitivity through differential detection, as $\Delta \Psi_{diff}$ in (\ref{eq:diff}) isolates the chiral contribution while canceling common-mode perturbations.

Thus, $\Psi_{OAM}$ inherits the concentration dependence through $\theta_{OAM}(C)=\Psi_{OAM}/\ell$ (via $\Delta n_{CB}$), providing a direct, quantitative link between glucose-induced optical activity and the measured twist of the OAM phase. We emphasise that the preserved observable is not the full optical phase but its azimuthal gradient, which survives as a statistical invariant under multiple scattering.

Further experimental validation of the chiral OAM phase memory across scattering strength, molecular chirality, and OAM charge is provided in the Supplementary Information.

\section*{Discussion}
These results establish chiral phase memory as a distinct phenomenon within the broader family of optical memory effects. Unlike spatial or angular memory, which preserve correlations through statistical averaging, chiral phase memory encodes molecular information into a topological degree of freedom that remains quantitatively accessible. This distinction has immediate implications: any chiroptical measurement currently limited by scattering, circular dichroism spectroscopy, optical rotation detection, and vibrational optical activity may benefit from OAM-based encoding.

Conventional circular dichroism (CD) and optical rotatory dispersion (ORD) measurements require ballistic or quasi-ballistic photon regimes, fundamentally limiting their penetration depth in biological tissue. The chiral phase memory demonstrated here operates in a regime ($z/l^* \approx 10$) where conventional polarimetric signals are completely destroyed, representing a qualitative extension of chiroptical detection capabilities. The key distinction is that chirality is encoded not in polarization state, which scattering randomises, but in the azimuthal gradient of a topological phase structure, which scattering preserves. Moreover, the ability to optimize sensitivity by selecting topological charge represents a unique advantage of structured light over conventional polarimetric approaches, which lack an analogous tuneable geometric parameter.

Detecting enantiomer-specific optical activity through turbid biological tissue has direct implications for non-invasive glucose monitoring and, more broadly, for pharmaceutical enantiomer detection where chiral purity is critical. Translation to \textit{in vivo} measurement will require addressing tissue heterogeneity, motion artifacts, and the development of compact OAM generation and detection systems.

Beyond chirality, our results suggest that topological phase observables can serve as robust carriers of weak light-matter interaction signatures more generally. The differential measurement strategy demonstrated here, comparing conjugate topological charges to isolate symmetry-breaking contributions may extend to other phenomena coupling differentially to OAM handedness, including magneto-optical effects and certain nonlinear processes. The $10^{-6}$ refractive index sensitivity achieved here approaches shot-noise limits for classical interferometry, raising the question of whether quantum-enhanced OAM states could push sensitivity further in scattering environments. The preservation demonstrated here persists to at least $z/l^* \approx 10$; characterizing the ultimate scattering threshold at which topological protection degrades remains an important direction for future investigation.

By coupling molecular chirality to beam topology, we establish a measurement paradigm in which the information carrier, the azimuthal phase gradient, is intrinsically protected by the topological structure of the light field, enabling chiroptical spectroscopy beyond the ballistic-photon regime.

\section*{Methods}

\subsection*{Experimental setup / Optical configuration}
A modified Mach--Zehnder interferometer configuration was employed for OAM phase measurements (Fig.~\ref{fig:glucose1}-a). A coherent Gaussian beam from a $640~nm$ laser diode ($40~mW$, BioRay, Coherent, USA; coherence length $>20~cm$) was coupled into a single-mode optical fiber (P1-630A-FC-1, Thorlabs) and collimated (F280FC-B, Thorlabs) to produce a beam with $1.6~mm$ waist diameter. A polarizer established horizontal linear polarization before a polarizing beam splitter divided the beam into sample and reference arms.

In the sample arm, a phase-only spatial light modulator (PLUTO-2-NIR-011, Holoeye, Germany), operating in reflection, generated LG beams with topological charges $\ell = \pm 3$ and $\ell = \pm 5$ via forked diffraction gratings. The first-order diffracted beam was spatially filtered through a pinhole and recollimated ($L3$, $L4$: $f = 45~mm$, Thorlabs) before traversing the sample. The reference Gaussian beam was expanded ($L1$: $f = 30~mm$; $L2$: $f = 70~mm$) and its polarization adjusted via a half-wave plate before recombination at a beam splitter.

Interference patterns were recorded using a CMOS camera (DCC3240M, $1280 \times 1024$ pixels, Thorlabs) with a $10\times$ objective (Nikon). The beam waist diameters at the detector plane were $2.7~mm$ ($LG^{3}_{0}$) and $3.0~mm$ ($LG^{5}_{0}$). The setup enabled both on-axis (petal rotation tracking) and off-axis (Fourier-transform phase retrieval) interferometric measurements~\cite{Meglinski}. For glucose measurements, 10 consecutive frames were acquired at controlled temperature ($21 \pm 0.5^{\circ}C$) and averaged to improve signal-to-noise ratio.

LG beams with $\ell = \pm 3$ were used for initial calibration measurements (Fig.~\ref{Phase}), while differential glucose sensing employed $\ell = \pm 5$ to maximize chiral phase sensitivity through the $|\ell|$ dependence in Equation (\ref{eq:diff}).

\subsection*{Phase Retrieval Methodology}

We employ off-axis vortex interferometry with Fourier-transform phase retrieval~\cite{Meglinski} to measure glucose-induced OAM phase shifts. The LG beam carrying OAM is superimposed with a reference plane wave at a slight angle, producing spiral interferograms where azimuthal rotation, rather than perpendicular fringe displacement, directly encodes optical path differences. 
This unique property of vortex beams enables single-shot phase measurement without temporal phase-shifting algorithms, with the $2\pi$ periodicity of the helical phase providing intrinsic calibration.

Phase retrieval was performed using established off-axis vortex interferometry and Fourier-domain reconstruction methods for structured light fields, following the inverse spiral-transform formalism described in Rodríguez-Zurita \textit{et al}. and systematically reviewed by Dong \textit{et al}.~\cite{Review}.

\subsection*{Decomposition of accumulated phase in chiral scattering media}
The detected phase of the transmitted LG beam can be decomposed into an intrinsic (geometric) contribution and a medium-induced contribution. This section introduces the phase terms used to interpret the measured azimuthal rotation and to distinguish the achiral background phase from the chiral (circular-birefringence) contribution.

At the detection plane, the total accumulated phase is written as the sum of a geometric contribution determined by the optical field structure and a medium-induced contribution arising from light–matter interaction:
\begin{equation}
\Psi(\rho, \varphi, z; C, \ell) = \Psi_{geometric}(\rho, \varphi, z) + \Psi_{medium}(C, \ell). 
\label{Psitot}
\end{equation}
where the geometric phase includes the intrinsic LG beam structure:
\begin{equation}
\Psi_{geometric}(\rho,\phi,z) = -\frac{k\rho^2 z}{2(z^2+z_R^2)} - \ell\phi - kz + G(z),
\label{Psi}
\end{equation}
with the terms representing radial wavefront curvature, the helical phase carrying the topological charge, longitudinal propagation phase, and the Gouy phase $G(z)=(2p+|\ell|+1)arctan(z/z_R)$.

The medium-induced phase is separated into achiral and chiral components:
\begin{equation}
\Psi_{medium}(C, \ell) = \Psi_{achiral}(C) + \Psi_{chiral}(C, \ell), 
\label{Psimedium}
\end{equation}
with:
\begin{equation}
\Psi_{achiral}(C) = k \, \Delta n_{achiral}(C) \, L_{eff}, 
\label{Psiachiral}
\end{equation}
\begin{equation}
\Psi_{chiral}(C, \ell) = k \, \text{sign}(\ell) \, \Delta n_{CB}(C) \, L_{spiral}(\rho) \cdot |\ell|. 
\label{Psichiral}
\end{equation}
The achiral term (\ref{Psiachiral}) depends on an effective optical path length $L_{eff}$ that includes contributions from bulk propagation and scattering-induced path elongation, but is independent of topological charge. The chiral term (\ref{Psichiral}) depends on the spiral trajectory length $L_{spiral}(\rho)$, which varies with radial position, and critically carries the $\ell$-dependence that enables differential sensing.

For LG beams, the effective spiral interaction length $L_{\text{spiral}}$ arises from the spatial dispersion of Poynting vector trajectories in LG beams~\cite{Berry2008,Bliokh2013}, defined as:
\begin{equation}
\begin{aligned}
&   L_{spiral} = L(r_0,\varphi_0,\zeta_s) = \int_0^{\zeta_s} \sqrt{ \left(\frac{dr}{d\zeta}\right)^2 + r^2\left(\frac{d\varphi}{d\zeta}\right)^2 } d\zeta, \\
&    r(\zeta) = r_0\sqrt{1+4\zeta^2}, \quad \varphi(\zeta)=\frac{\ell}{2 r_0^2}\arctan(2\zeta) + \varphi_0,
\end{aligned}
\label{eq:LGtrj}
\end{equation}
where $L(r_0,\varphi_0,\zeta_s)$ defines the Poynting vector trajectory length between $z=0$ and any non-zero $z_s$, $r(\zeta), \varphi(\zeta)$ are cylindrical coordinates of the single point on trajectory, $r_0, \varphi_0$ are coordinates of the trajectory starting point at $z=0$, ${\zeta=\dfrac{z}{k{w}^2(0)}}, r=\dfrac{\rho}{{w}(0)}$; derived using Poynting vector direction ($\mathbf{p}=\{p_\rho, p_\phi, p_z\}$)~\cite{Allen1999-uc}:
\begin{equation}
\begin{aligned}
& p_{r}= \frac{\omega k r z}{\left(z_{R}^{2}+z^{2}\right)} \lvert LG^\ell_p(r,\phi,z)
\rvert^{2}, \\
& p_{\phi}=\frac{\omega l}{r} \lvert LG^\ell_p(r,\phi,z)\rvert^{2}, \\
& p_{z}= \omega k \lvert LG^\ell_p(r,\phi,z)\rvert^{2} \text {, }
\end{aligned}
    \label{eq:LGbeamDirections}
\end{equation}
where $\omega$ is the angular frequency of the light.

\textit{Note}: In the main Results, we focus on the experimentally observable azimuthal rotation and differential cancellation; Eqs. (\ref{Psitot})–(\ref{Psichiral}) are provided here for completeness and for parameter definitions used in the analysis.

\subsection*{Relation between azimuthal OAM phase rotation and optical activity}

The glucose-induced chiral contribution to the OAM phase $\Psi_{glucose}$ (equivalent to $\Psi_{chiral}$ in (\ref{Psichiral})) arises from circular birefringence accumulated along the effective spiral trajectories of the LG beam (\ref{eq:LGtrj}). For an OAM mode of topological charge $\ell$, the chiral phase accumulation can be written as
\begin{equation}
\Psi_{glucose}(\rho,C) = \ell k \Delta n_{CB}(C) L_{spiral}(\rho).
\label{eq:Psdu-gluc}
\end{equation}
where $\Delta n_{CB}(C) = n_L(C) - n_R(C)$ is the circular birefringence of the glucose solution and $L_{spiral}(\rho)$ is the effective spiral interaction length at radial coordinate $\rho$.
The corresponding azimuthal rotation angle of the helical phase is defined as
\begin{equation}
\theta_{OAM}(C) = \frac{\Psi_{glucose}}{\ell} = k \Delta n_{CB}(C) L_{spiral}(\rho),
\label{eq:thetaOAM}
\end{equation}
where circular birefringence $\Delta n_{CB}(C)$ is related to the specific rotation $[\alpha]$, as commonly defined in polarimetry~\cite{Barron_2004}, via $\Delta n_{CB} = \frac{\lambda}{\pi}[\alpha]C$, although all analysis in this study is performed directly in terms of $\Delta n_{CB}(C)$. Unlike conventional polarimetry, where the measured rotation corresponds to half of the relative phase delay between left- and right-circular polarization components, the azimuthal phase rotation of an OAM beam reflects the full circular-birefringence-induced phase accumulation along the spiral energy-flow trajectory.

\subsection*{Samples} 

To show the possibility of glucose sensing using LG light carrying OAM the glucose-water solutions with 50, 70, 90, 110, 130 and $150~mg/dl$ concentrations were prepared using $D-(+)$-glucose powder (Sigma-Aldrich, USA). Each solution was inserted into the sample cuvette. In addition, a series of 10 consecutive frames of the interference pattern in both the on-axis and the off-axis regimes were captured at controlled temperature ($21^{\circ}\text{C}$) for further average and analysis. To demonstrate the possibility of glucose sensing, the \textit{ex vivo} porcine skin tissue sample and/or low-scattering ($z/l^{*} = 2$, $z = 8~mm$) phantom medium, both placed in front of the cuvette containing the glucose solution in water (as presented in Fig.~\ref{fig:glucose1}-(a)) were used. The detailed procedure for the preparation of the phantom is presented in~\cite{wrobel2015measurements}.

\section*{Data availability}
The data supporting the findings of this study are available from the corresponding author upon reasonable request.

\section*{Code availability}
The custom code used for Fourier-domain phase retrieval and azimuthal phase analysis is available from the corresponding author upon reasonable request.

\section*{Acknowledgements}
The authors are grateful to Dr. Ivan Lopushenko (University of Oulu, Finland) for useful discussions and valuable input during the study and manuscript preparation. This work was supported by EU Horizon Europe EIC Pathfinder Open Research and Innovation Programme, OPTIPATH project, Grant Agreement No. 101185769, and partially supported by COST Action CA21159 -- Understanding interaction light -- biological surfaces: possibility for new electronic materials and devices (PhoBioS) and COST Action CA23125 -- The mETamaterial foRmalism approach to recognize cAncer (TETRA) funded by the COST (European Cooperation in Science and Technology).

\section*{Author contributions statement} AS, AB, IM designed the concept of the study. AS performed measurements and processed data and implemented the software for data processing and contributed to data analysis. IM wrote the manuscript. AB contributed to data analysis, discussions, and reviewed the manuscript. All authors have accepted responsibility for the entire content of this manuscript and consented to its submission to the journal, reviewed all the results and approved the final version of the manuscript. 

\section*{Additional information}
The authors have declared that no competing interest exists.

\bibliography{sample}

\end{document}